\documentclass[9pt]{IEEEtran}
\usepackage{graphicx}
\usepackage{tikz}
\usepackage{amsmath}
\usepackage{amssymb}
\usepackage{times}
\usepackage{subfigure}
\usepackage{color}

\begin{document}

\sloppy

\pagestyle{plain}

\title{SAPA: Self-Aware Polymorphic Architecture} 

\author{Michel A. Kinsy, Mihailo Isakov, Alan Ehret, Donato Kava\\
Adaptive and Secure Computing Systems Laboratory \\
Boston University, Boston, USA
\vspace{-0.4in}}

\maketitle
\begin{abstract}
  In this work, we introduce a Self-Aware Polymorphic Architecture (SAPA) design approach to support emerging context-aware applications and mitigate the programming challenges caused by the ever-increasing complexity and heterogeneity of high performance computing systems. Through the SAPA design, we examined the salient software-hardware features of adaptive computing systems that allow for (1) the dynamic allocation of computing resources depending on program needs (e.g., the amount of parallelism in the program) and (2) automatic approximation to meet program and system goals (e.g., execution time budget, power constraints and computation resiliency) without the programming complexity of current multicore and many-core systems. The proposed adaptive computer architecture framework applies machine learning algorithms and control theory techniques to the application execution based on information collected about the system runtime performance trade-offs. It has heterogeneous reconfigurable cores with fast hardware-level migration capability, self-organizing memory structures and hierarchies, an adaptive application-aware network-on-chip, and a built-in hardware layer for dynamic, autonomous resource management. Our prototyped architecture performs extremely well on a large pool of applications.  
\end{abstract}

\vspace{-0.1in}
\section{Introduction}
\label{sec:intro}
\begin{figure}[htb]
\vspace{-0.1in}
\begin{center}
\includegraphics[width=3.0in]{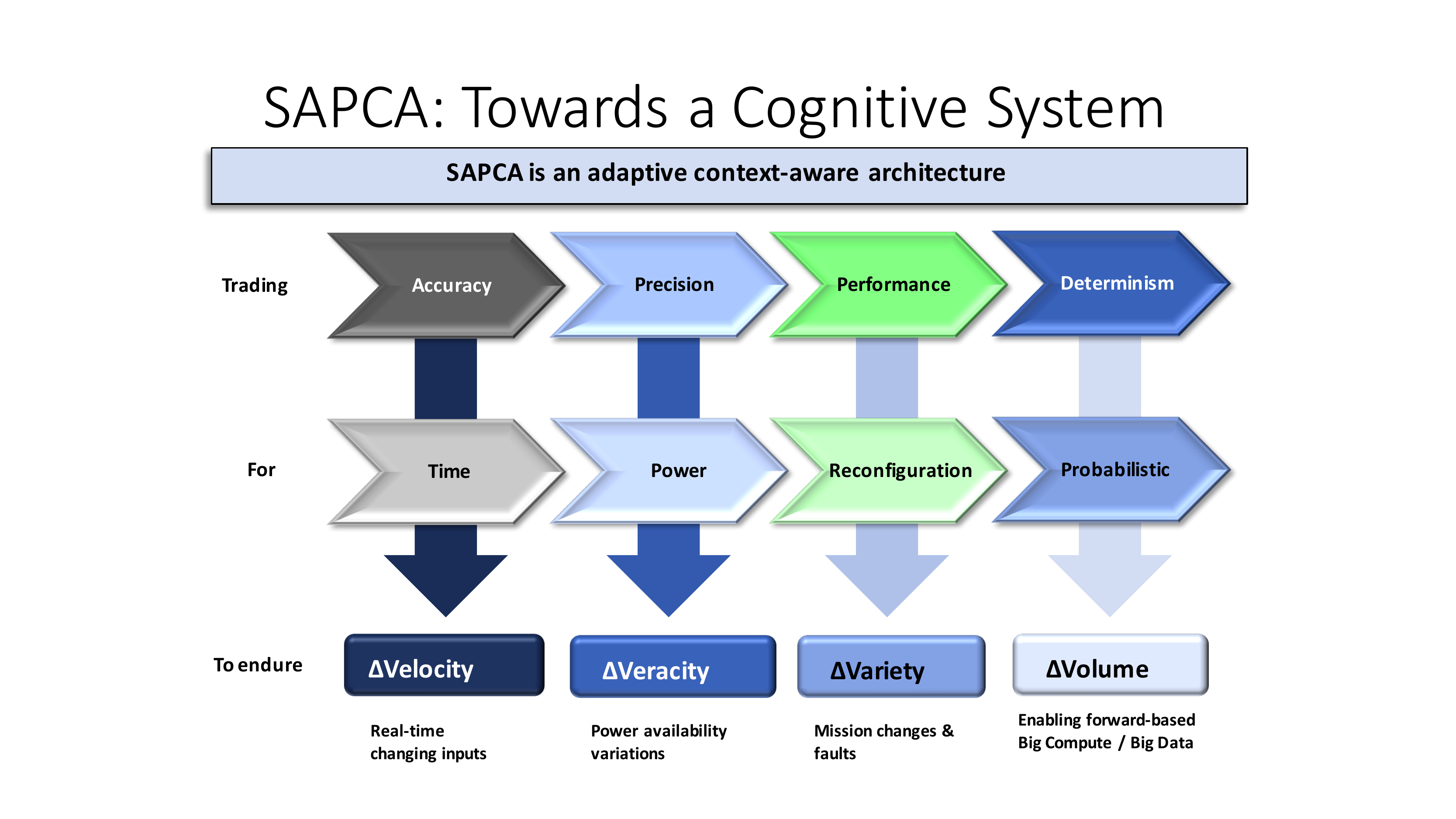}
\vspace{-0.15in}
\caption{Salient desired adaptation range in new computing systems.}
\label{fig:adaptation}
\end{center}
\vspace{-0.3in}
\end{figure}

The current design approach in multicore or many-core computer systems presents application programmers 
with a great deal of challenges due to their ever-increasing complexity~\cite{Kephart}. Unlike frequency-scaling where 
performance is increased equally across the board, core-scaling pushes the burden of harnessing this additional 
processing power on the application or software programmer~\cite{Laddaga}. To make optimal use of the system 
components, programmers must first learn about system parameters and how to best leverage them for a given 
application. This approach requires time, effort, and often leads to suboptimal application performance in terms of 
execution time or power. Furthermore, it has become evident that better data processing capabilities are needed to extract meaningful insights from 
large-scale data~\cite{Kephart}. Exhaustive or deterministic executions of 
these extreme-scale applications are increasingly too expensive under current systems in terms of power and 
execution time. Figure~\ref{fig:adaptation} depicts the desired trade-off choices to adequately target new 
emerging applications. 

\section{Related Work}
\label{sec:related}
Acknowledging the difficulty of managing the complexity of future computing systems, Kephart et al. \cite{Kephart} suggested that one of the viable approaches to solving the problem lies in the design and development of self-adaptive computing systems. Among others, Albonesi et al. \cite{Albonesi} showed how adaptive processing can improve microprocessor energy efficiency by dynamically tuning major microprocessor resources, e.g., caches, hardware queues, during execution to better match varying application needs. They further highlighted that adaptive systems require few additional transistors. Recently, Hoffmann et al. \cite{Hoffmann} presented the Self-aware Computing (SEEC) model where application goals could be implemented to help guide the runtime system execution. The proposed architecture framework builds on these research efforts and insights.

\vspace{-0.1in}
\section{Self-Aware Polymorphic Architecture (SAPA) Design}
\label{sec:arch}

\begin{figure}[htb]
\vspace{-0.05in}
\begin{center}
\includegraphics[width=3.0in]{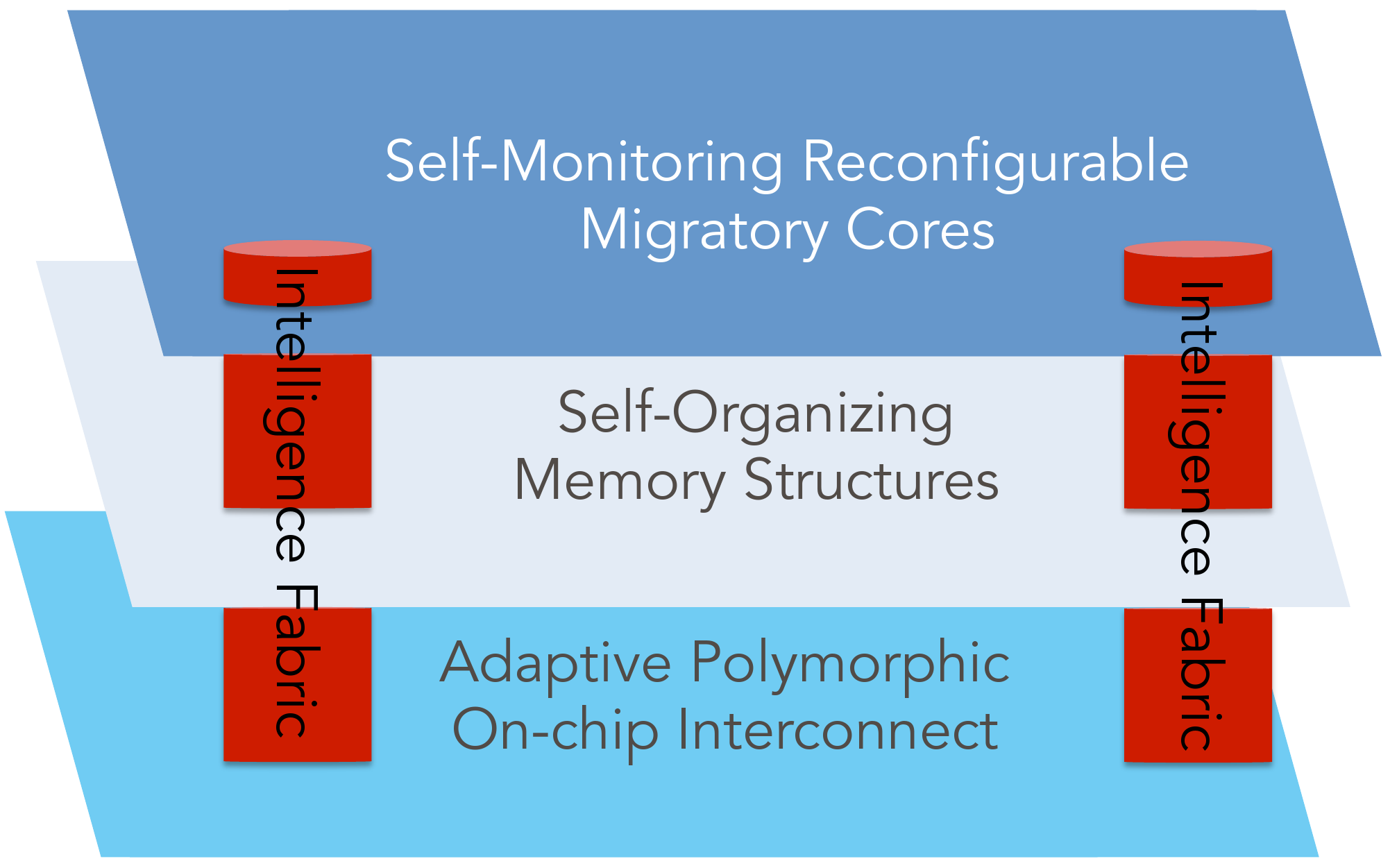}
\vspace{-0.1in}
\caption{Illustrative architectural stack of self-Aware Polymorphic Architectures (SAPA).}
\label{fig:stack}
\end{center}
\vspace{-0.15in}
\end{figure}

For a computing system to automatically and dynamically adapt to program execution constraints, goals and phases, it needs to sense hardware states (a way to gather runtime system information), monitor program execution phases, and make complex decisions based on built-in execution rules. 
SAPA extends the conventional multicore/many-core architecture three-layer design approach, consisting of processing elements, memory subsystems and the on-chip data communication logic with a fourth layer, i.e., the intelligence fabric (IF) or nervous system (NS) layer. This new distributed introspection and reconfiguration architectural layer is the central conceptual innovation of the SAPA design methodology and enables continuously autonomous adaptation for performance, reliability, and energy efficiency. Figure~\ref{fig:stack} illustrates the SAPA computing architecture stack. It has a fast hardware-level migration capability at the cores, self-organizing memory structures and hierarchies, an adaptive and quality-of-service aware network-on-chip, and the built-in NS layer for dynamic, autonomous resource management. 

The new set of architectural layers for the SAPA stack are: 
\begin{itemize}
\item \textbf{Self-Aware Polymorphic Execution Cores (SAPEC)}: processing elements (PEs) that dynamically adapt and optimize their execution behavior according to a set of high-level program goals. 
\item \textbf{Approximation-Aware Memory Organization Models (AMOM)}: smart, self-adjusting distributed memory structures that use hardware counters for online execution pattern learning and correctness estimation.
\item \textbf{Resilient Adaptive Intelligent Network-on-Chip (RAIN)}: routers that address runtime network traffic load imbalances and reliability problems in network-on-chip design in the presence of faults. 
\item \textbf{Dynamic Approximation Execution Manager (DAEM)}: distributed execution control software-hardware modules, which collect runtime system and application information and apply machine learning algorithms to achieve desired performance and resource utilization through adaptation and approximation. 
\end{itemize}

\begin{figure}[htb]
\vspace{-0.05in}
\begin{center}
\includegraphics[width=3.0in]{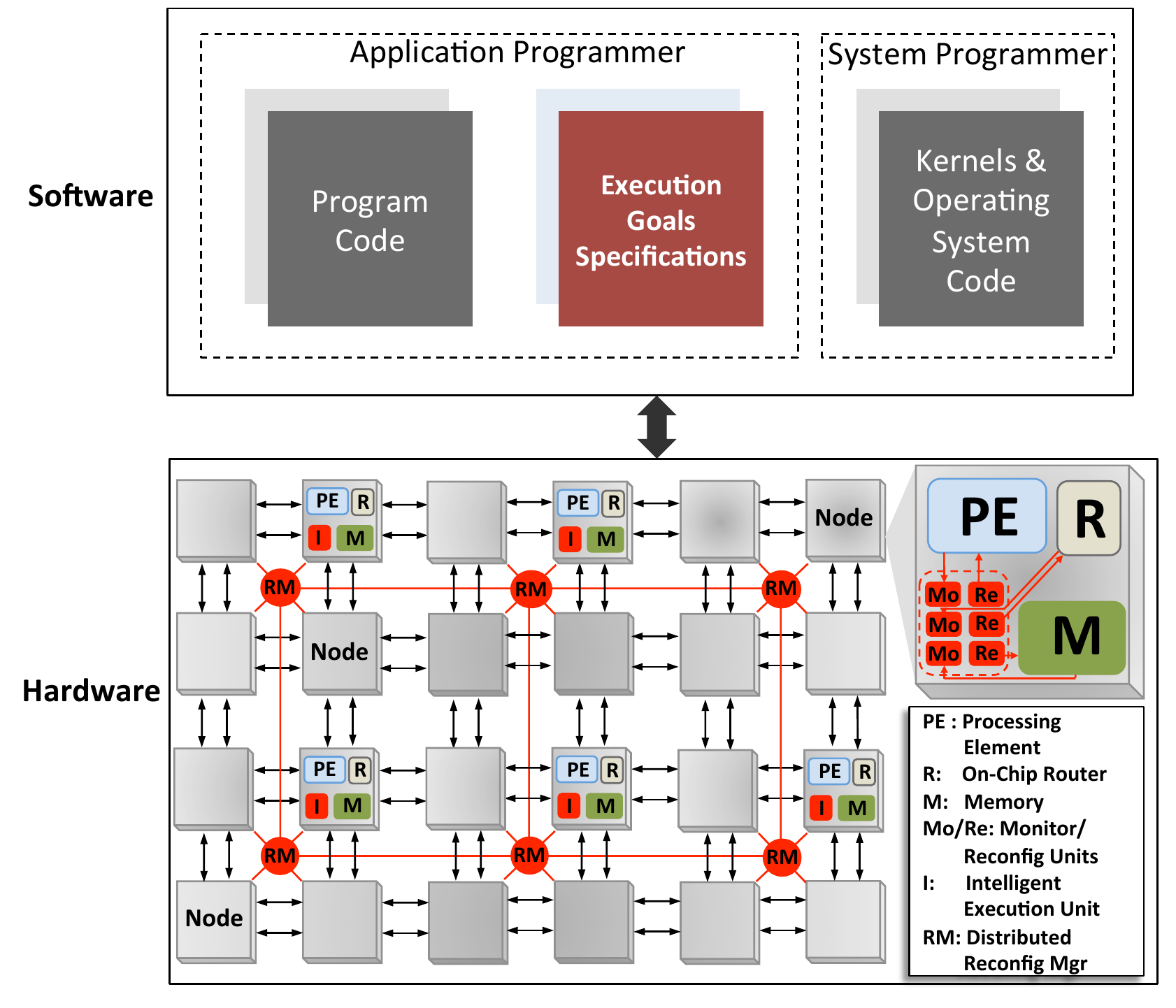}
\vspace{-0.1in}
\caption{Self-Aware Polymorphic Architecture (SAPA) system.}
\label{fig:sapa}
\end{center}
\vspace{-0.15in}
\end{figure}

Figure \ref{fig:sapa} shows a concrete implementation overview of the architecture: 1) processing elements, 2) a distributed memory subsystem, 3) an interconnection network, and 4) sensing, monitoring, and reconfiguration hardware components (the intelligent sensing and monitoring components are indicated in red). The novel intelligence layer, the \textbf{\textit{nervous system (NS)}}, is composed of the distributed \textbf{\textit{intelligent execution unit (I)}} and \textbf{\textit{reconfiguration manager (RM)}}. 

Given a program, its decomposition, and its performance targets (e.g., power, latency), the system dynamically allocates computational resources (PEs) depending on the amount of parallelism in the program, memory footprint, and data communication routing overhead. The dynamic mapping of the program can take different shapes depending on the number of processing elements (or maximum allowed simultaneously active processing elements due to power budget) on the architecture, current local network congestion, and data dependencies between tasks over time. The on-chip network of active processing elements expands and contracts dynamically during runtime depending on the program execution needs and performance-power targets. Other novel features of the architecture include (1) the use of neighbor cores for rehearsal of decisions, (2) having non-global cache coherence for scalability and adaptability, and (3) creating a framework for parent-child model of data replication. 

The following steps demonstrate program execution on the designed SAPA system.  
\noindent\textbf{\textit{Sample execution scenario}}:
\begin{enumerate}
\item The program and user constraint specifications, in the form of pragmas, are compiled; 
\item The compiler partitions the program and maps it to hardware resources; 
\item Program codes are generated alongside monitoring codes with customized reconfiguration rules; 
\item New system libraries are added to the linkage, loading and image stages; 
\item Images are loaded to main memories and execution starts; 
\item Monitoring counters begin capturing processing element, memory and network events;  
\item Violation of one or more constraints triggers an adaptation interrupt; 
\item The distributed reconfiguration manager applies analysis and reconfiguration rules including resource re-allocations, task rehearsal and task migrations; 
\item Execution resumes. 
\end{enumerate}

\begin{figure}[htb]
\vspace{-0.05in}
\begin{center}
\includegraphics[width=2.75in]{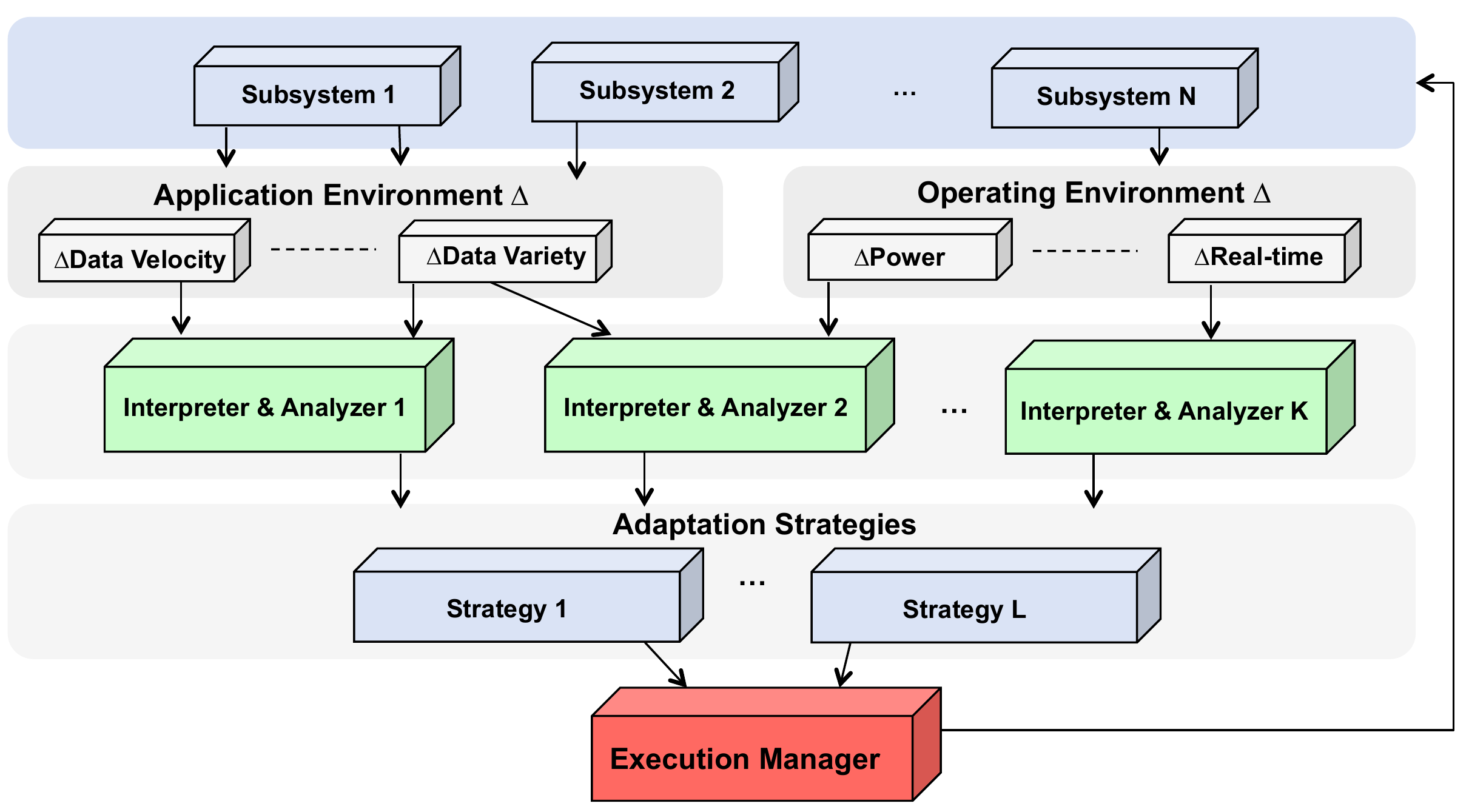}
\vspace{-0.1in}
\caption{Conceptual view of the autonomous adaptive execution model.}
\label{fig:concept}
\end{center}
\vspace{-0.2in}
\end{figure}

Figure \ref{fig:concept} is a static depiction of a dynamic SAPA execution. The \textit{Interpreters} synthesize hardware instructions from program goals specifications. The \textit{Analyzers} collect sensing information and perform system state analyses. The \textit{Reconfiguration Manager (RM)} applies techniques from artificial intelligence to make decisions concerning reconfigurations. These decisions are communicated to the distributed \textit{intelligent execution unit (I)} within the nodes to reconfigure the PE, router and local memory. Self-awareness, in the context of this work, refers to the computer system's ability to monitor, collect and analyze data about its own state of execution.

\vspace{-0.1in}
\section{Evaluations}
\label{sec:results}
\begin{figure}[htbp]
\begin{center}
\includegraphics[width=3.25in]{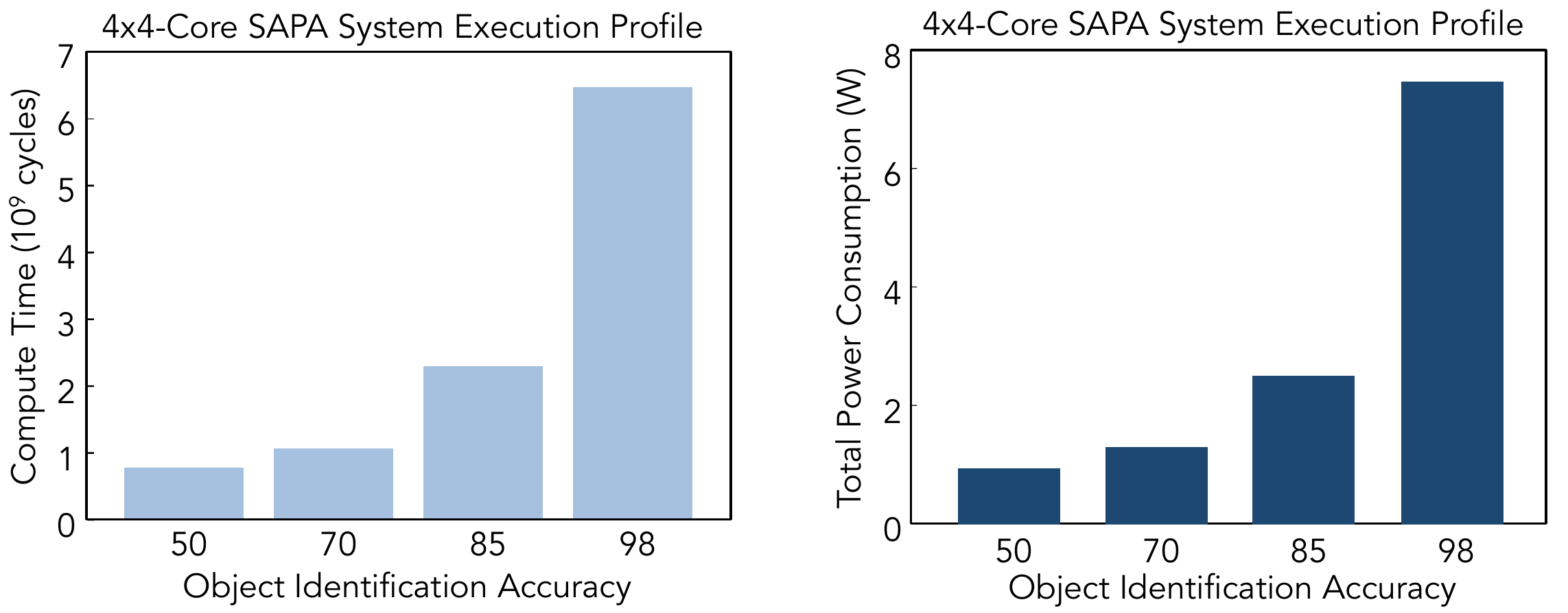}
\vspace{-0.1in}
\caption{Fast object recognition application execution.}
\label{fig:runs}
\end{center}
\vspace{-0.3in}
\end{figure}

To evaluate the proposed SAPA design, we hand-annotated Betke {\em et al.}~\cite{Betke} {\em Fast Object Recognition in Noisy 
Images Using Simulated Annealing} algorithm and ran it using the Heracles \cite{heracles} design tool with some architectural changes. 
Figure~\ref{fig:runs} shows how these types of iterative algorithms lend themselves well to user-defined or context-aware \textit{execution time to accuracy} or \textit{power to accuracy} trade-offs.
For this application, moving from 85\% to 98\% matching confidence level triples the power 
consumption and the compute time. While in some cases a 98\% matching confidence level may be required, the extra compute cost 
may not be justifiable under other operating circumstances. 

\vspace{-0.05in}

\bibliographystyle{IEEEtran}
\bibliography{paper}


\end{document}